\pdfoutput=1 
\documentclass{JINST}
\usepackage{graphicx}
\usepackage{caption}
\usepackage{subcaption}
\usepackage{gensymb}

\title{Position Reconstruction of Bubble Formation in Liquid Nitrogen using Piezoelectric Sensors}

\author{Brian Lenardo $^{a,b}$,
Yin Li$^a$,
Aaron Manalaysay$^a$,
James Morad$^a$,
Chrisman Payne$^a$,
Scott Stephenson$^a$,
Matthew Szydagis$^c$ and
Mani Tripathi$^a$\\
\llap{$^a$}University of California, Davis\\
One Shields Avenue, Davis, CA 95616\\
\llap{$^b$}Lawrence Livermore National Laboratory,\\
7000 East Avenue, Livermore, CA 94550 \\
\llap{$^c$}University at Albany, SUNY\\
1400 Washington Ave, Albany, NY 12222\\
}

\abstract{Cryogenic liquids, particularly liquid xenon and argon, are of interest as detector media for experiments in nuclear and particle physics. Here we present a new detector diagnostic technique using piezoelectric sensors to detect bubbling of the liquid. Bubbling can indicate locations of excess heat dissipation e.g., in immersed electronics. They can also interfere with normal event evolution by scattering of light or by interrupting the drift of ionization charge.  In our test apparatus, four sensors are placed in the vacuum space of a double-walled dewar of liquid nitrogen and used to detect and locate a source of bubbling inside the liquid volume. Utilizing the differences in transmitted frequencies through the different media present in the experiment, we find that sound traveling in a direct path from the source to the sensor can be isolated with appropriate filtering. The location of the source is then reconstructed using the time difference of arrivals (TDOA) information.  The reconstruction algorithm is shown to have a 95.8\% convergence rate and reconstructed positions are self-consistent to an average $\pm0.5$~cm around the mean in $x$, $y$, and $z$. Systematic effects are observed to cause errors in reconstruction when bubbles occur very close to the surfaces of the liquid volume. }


\begin{document}

\section{Introduction}\label{sec:Intro}
\paragraph{}Detectors constructed of large volumes of cryogenic liquids, particularly argon or xenon, are widely used in rare event searches in nuclear and particle physics \cite{LUX,LZ,EXO,XENON,DARKSIDE,PANDAX}. These experiments measure scintillation and/or ionization produced by impinging radiation. One particular technique is the dual-phase time projection chamber (TPC).  The TPC consists of a large volume of liquid that acts as the target with a small layer of gas at the top of the detector to allow for amplification of the ionization signal through electroluminescence. To maintain the two phases in equilibrium, the liquid must be held very near its boiling point. In such an environment, bubbles have been observed to form either spontaneously or in regions where there is some active heat dissipation from components such as readout electronics.

Bubbles can interfere with the normal operation of detectors in several ways. First, the liquid-gas interface at the surface of the bubble can scatter scintillation and electroluminescence light, interfering with readout and position reconstruction of radiation events in the detector. Second, bubbles that make it to the surface of the liquid in dual-phase TPCs can change the uniformity and level of the interface, altering the electroluminescence signal from ionization electrons and interfering with energy and position reconstruction. Third, bubbles can disrupt the measurement of ionization charge if they fall in the path of drifting electrons. Finally, bubbles are a symptom of excess heat being generated in the detector, indicating a problem with electronic components or insulation. Therefore, it is desirable to be able to detect bubbles forming in the liquid and reconstruct their position in order to diagnose problems.

We have developed piezoelectric transducers that are capable of unambiguously detecting the sound created by the formation of bubbles in a volume of cryogenic liquid. In this work, we test these sensors in a cylindrical dewar of liquid nitrogen. We demonstrate their sensitivity to the sound of bubbles traveling through different media and the ability to reconstruct the 3D position of bubble formation using the time difference of arrival (TDOA) information.

\section{Experimental Apparatus}\label{sec:Exp}

\begin{figure}[tbp] 
\centering
\includegraphics[width=0.4\textwidth]{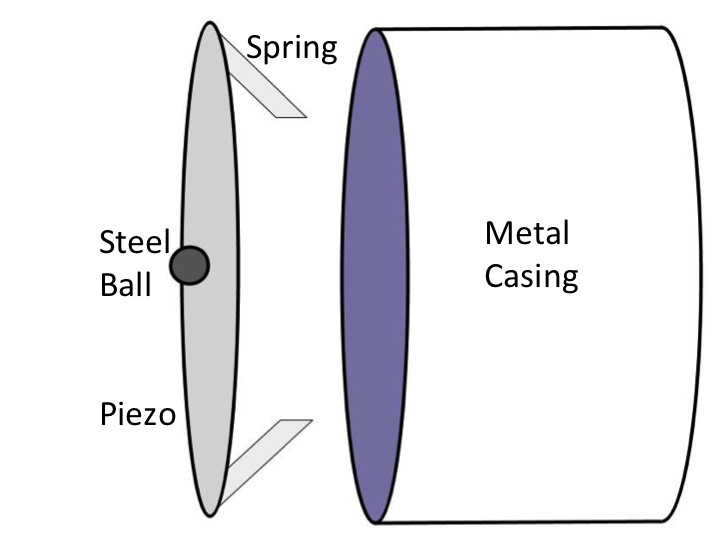}
\includegraphics[width=0.28\textwidth]{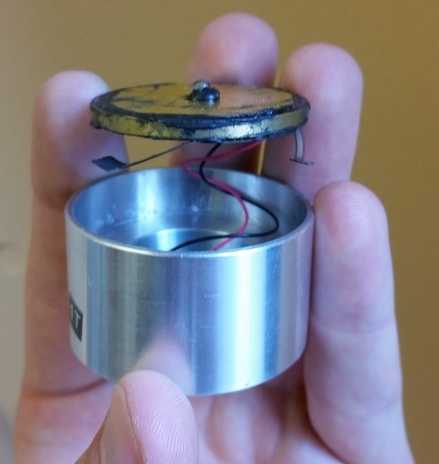}
\includegraphics[width=0.28\textwidth]{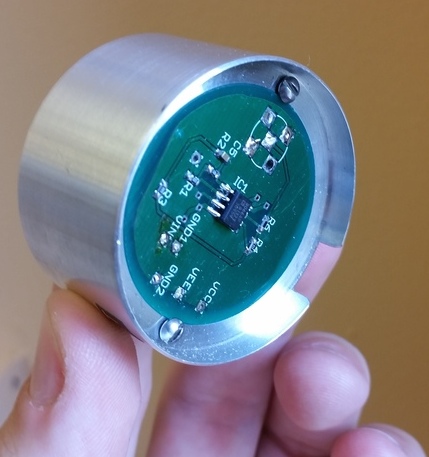}
\caption{Left: A schematic diagram of a sensor assembly; Middle: A photograph of the front side of the sensor showing the piezo, the steel ball and spring; Right: A photograph of the back side of the assembly showing the amplification electronics.} 
\label{fig:SensorDiagram}
\end{figure}

\paragraph{}The piezoelectric transducers have been developed to increase response by maximizing flexure of the piezoelectric element. Figure \ref{fig:SensorDiagram} shows a schematic diagram and photographs of the assembly. A ball bearing, secured to the center of the element with cryogenic glue, translates vibration from the stainless steel wall to the sensor. A spring-damper system at the edge of the piezoelectric element provides stability and enables the assembly to absorb high frequency noise and increase signal-to-noise ratio without affecting signal quality. The spring is a modified design from a conventional leaf spring, and the mass damper is a 1/4\"-thick brass ring to partially absorb the energy of high frequency oscillations. The leaf spring, brass ring and piezoelectric element are joined with cryogenically glue.  This design provides a higher relative displacement of the piezoelectric crystals, which results in higher signal amplitude than a piezoelectric element directly contacting the dewar.  An amplification circuit provides a gain of 215 to the transducer signal. High speed, low noise AD8066 JFET preamplifiers are used in the circuit design. 

\begin{figure}[tbp] 
\centering
\includegraphics[width=0.7\textwidth]{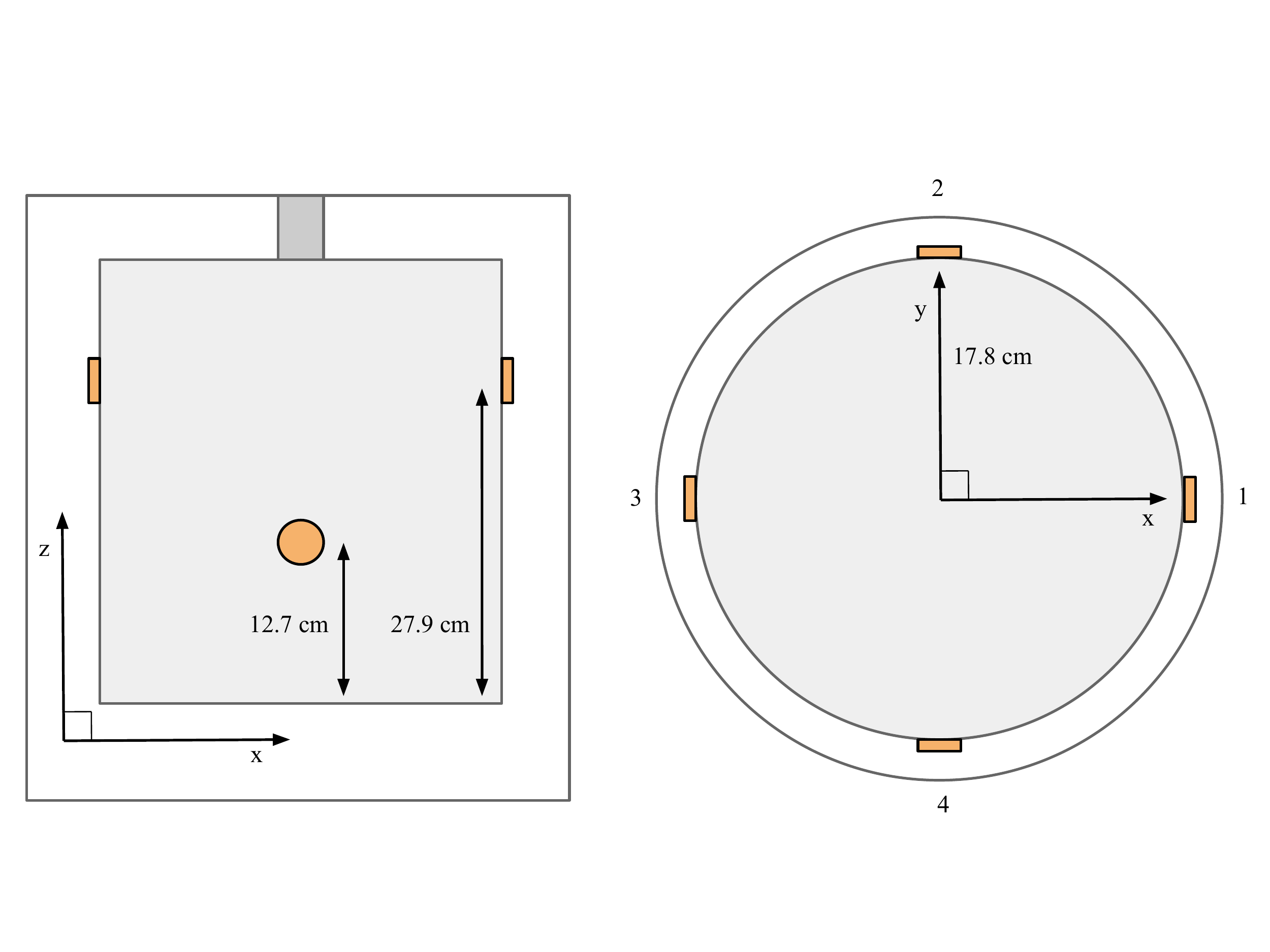}
\caption{Diagrams of the test dewar.  Sensors are shown in orange, and the inner volume containing liquid nitrogen in gray. Bubbles can be generated throughout the volume by means of a pulsed resistor attached to the end of a flexible, rotatable Garolite arm.}
\label{fig:DewarDiagram}
\end{figure}

To model a cryogenic liquid detector, we use a cylindrical stainless steel dewar filled with liquid nitrogen (LN), sketched in Figure \ref{fig:DewarDiagram}.  The the inner space has a volume of approximately 30 liters.  Four sensors are mounted and held against the surface using steel wire tensioned by springs. The ball bearing in the center of each piezo element is acoustically coupled to the dewar with a small amount of wax.  The sensors are mounted at $90\degree$ angles and at two different heights (centered at $z = 12.7$~cm and $z = 27.9$~cm) to reduce degeneracies in the position reconstruction.  Sensors that are $180\degree$ from each other are coplanar in $z$.

Bubbles can be generated throughout the dewar using a 100 $\Omega$ resistor connected to the end of a rotatable Garolite arm. Voltage pulses are sent to the resistor using a HP 8013b pulse generator. The minimum energy needed to create a bubble is found by raising the height and width of the supplied pulses until bubbles are observed, then decreasing the width until no bubbles are observed.  The final pulses used in this experiment have a height of 6.56~V and a width of 14.12~ms, resulting in 6.07~mJ delivered to the resistor with each pulse.  The pulses are supplied at a frequency of 1~Hz, which is observed to be long enough for the sound of the bubble to die away completely before the next pulse.  Sound signals are acquired by a Tektronix DPO3054 digital oscilloscope.  The piezoelectric sensors are fed into the four channels, digitized, and saved to an external flash drive for later analysis. Events that we use in the analyses below are triggered on the pulses sent to the resistor. Each waveform is sampled at 5~MHz and lasts 2 ms (10000 samples).

The data used in this analysis were taken on two separate days in April 2015 and June 2015. The first dataset consists of bubbles generated in the center of the dewar at different positions along the $z$-axis.  These data are used to study sound transmission in the experiment. The second dataset includes bubbles generated in a variety of positions throughout the volume, and is used in the position reconstruction analysis.

\section{Data Analysis}\label{sec:Ana}

\subsection{Sound transmission through different media}
\label{subsec:trans}

\begin{figure}[tbp]
    \centering
    \begin{subfigure}[b]{0.49\textwidth}
        \includegraphics[width=\textwidth]{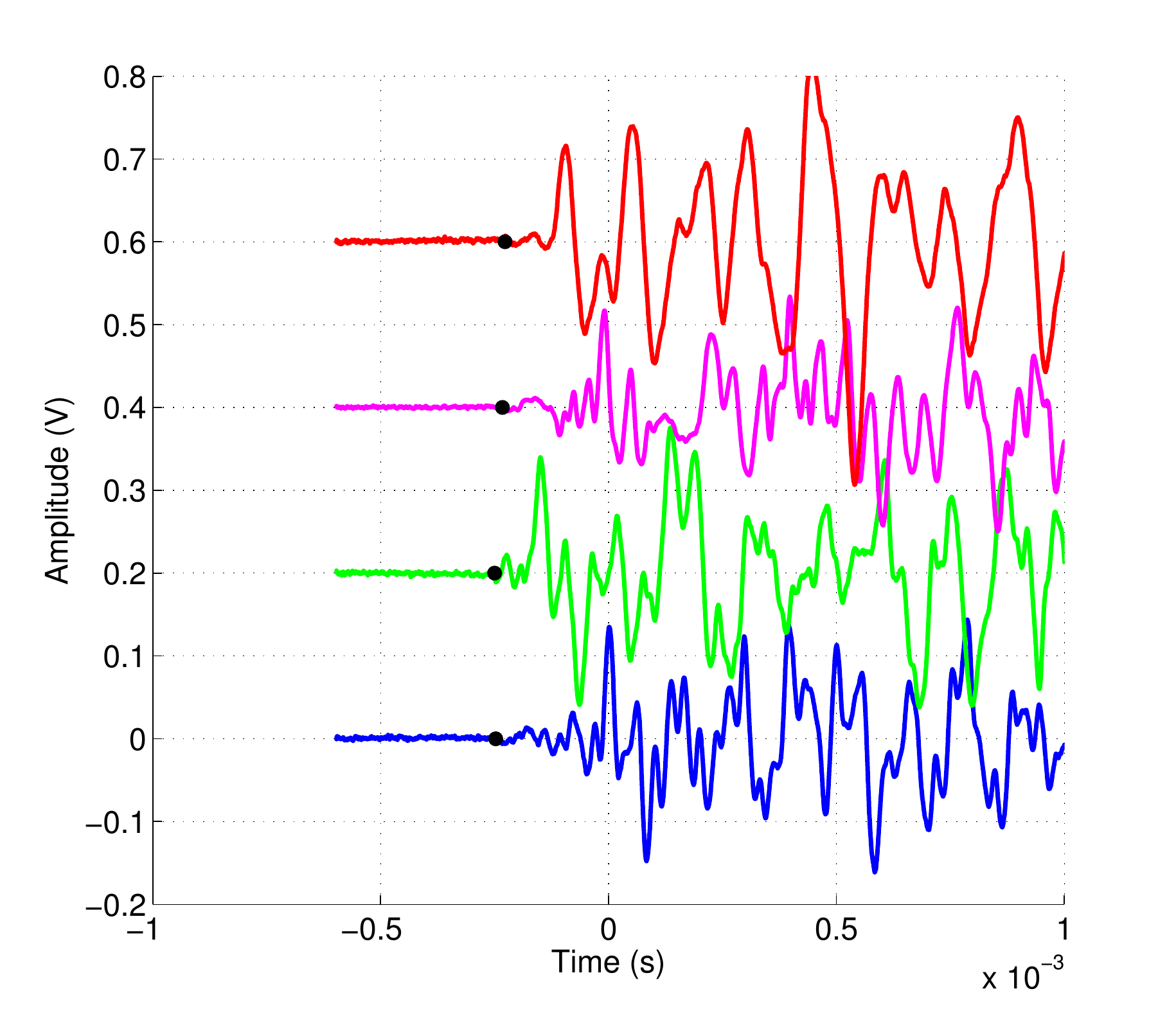}
        \caption{Bubble at center, z = 0 cm}
        \label{fig:BottomCenter}
    \end{subfigure}
    ~ 
    \begin{subfigure}[b]{0.49\textwidth}
        \includegraphics[width=\textwidth]{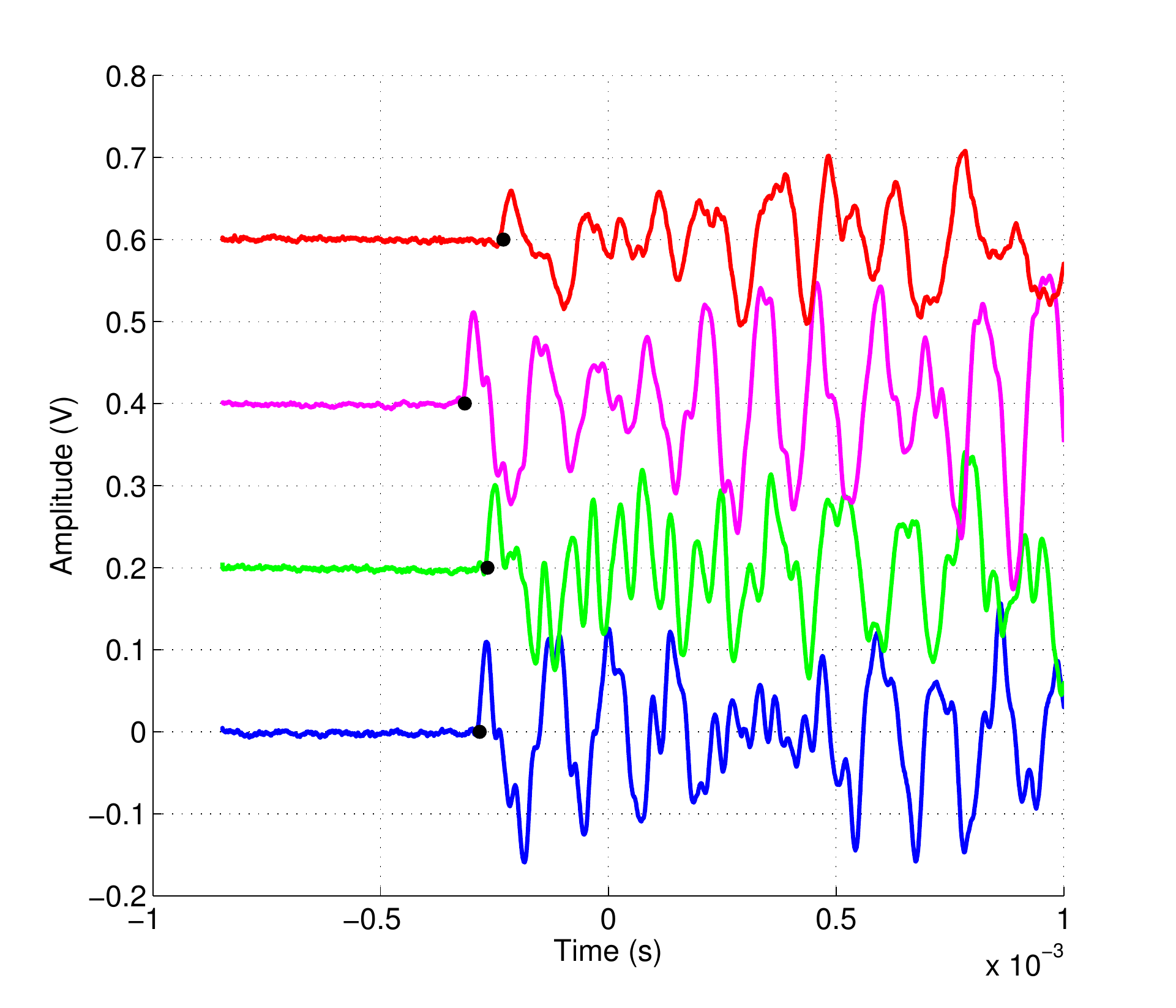}
        \caption{Bubble at center, z = 26.5 cm}
        \label{fig:TopCenter}
    \end{subfigure}
        \caption{Digitized sound waves of bubbles generated by the resistor in the center of the dewar.  Black points are sound arrival times reconstructed by setting a polarization-independent threshold of 10\% of the noise level in the raw signal. The difference between arrival times can be used to calculate bubble position. Low amplitude activity is visible in all channels at the beginning of the event at the bottom of the dewar. }
        \label{fig:TransmissionWaveforms}
\end{figure}

\paragraph{}The transmission of sound through the steel walls of the dewar is studied by comparing data taken at the bottom center of the dewar and the top center. Data taken at the bottom center of the dewar is very close to a steel surface and maximally distant from the sensors. Due to the much higher speed of sound in steel, the fastest path of travel is through the floor and then the wall of the container. Therefore, we expect transmission of sound through the steel to be visible in these traces at the beginning of the event.  In contrast, bubbles at the top center are maximally distant from any steel surfaces, so we expect the dominant contribution to sound arrival times to be due to transmission through the liquid nitrogen. In general, sound generated along the $z$-axis is expected to arrive in coplanar channels at the same time, and pairs of channels within an event can be used to cross-check each other.

\begin{table}
\centering
\begin{tabular}[tbp]{c|c|c|c|c}
Position (cm) & Threshold setting &  $t_2-t_1$ ($\mu$s)& $t_4 - t_3$  ($\mu$s) & $\Delta\,t_{exp}$\\
\hline
(0, 0, 1) & Low & $-25\pm 27 $ &  $7 \pm 23$ & -30 \\
(0, 0, 1) & High & $-153.8 \pm 0.4$ & $-88.2 \pm 0.5$ & -123\\
(0, 0, 26.5) & Low \& High & $17.7 \pm 0.3$ & $82.3 \pm 0.1 $ &54.7\\ 
\end{tabular}
\caption{Time differences used in the sound transmission analysis. For bubbles in the center of the dewar, we expect $\Delta\,t_{12} \approx \Delta\,t_{43}$. The low threshold is used to detect sound traveling through the steel for events close to to the bottom. The high threshold measures arrival time of sound traveling through the liquid nitrogen. For events far away from walls, both thresholds measure the same time. }
\end{table}

Two example events from the April 2015 dataset are shown in Figure \ref{fig:TransmissionWaveforms}. To remove noise in each trace, we apply a box filter with a width of 20 samples. The time of arrival of the sound is defined with two different thresholds. In low-threshold operation, the arrival time is defined as the point where the wave crosses a threshold of $\pm4\,\sigma_{noise}$, where $\sigma_{noise}$ is the standard deviation of the first 1000 samples in the smoothed waveform. In high-threshold mode, the arrival time is the point when the signal surpasses 30\% of the maximum height of the signal. The arrival times found in the traces in low-threshold mode are shown in Figure \ref{fig:TransmissionWaveforms}  by the black markers.  A low-amplitude signal is observed in the leading part of the trace for data taken at $z = 1$cm, while no such signal is observed in the data taken at $z = 21$cm. To confirm that this signal is sound traveling through the steel, we average together the differences in arrival times between the higher channels (1 and 3) and the lower channels (2 and 4) for all events at each location. The results are shown in Table 1. Uncertainties are calculated as the standard deviation of the measurements across all events. 

For the low-threshold setting, we are consistent with the expected time difference due to sound traveling through the steel. When the threshold is raised, the time differences are of the correct order of magnitude when compared sound traveling through the liquid nitrogen. There is a systematic difference observed in the two pairs of channels, likely caused by the resistor not being directly on the z-axis.  We estimate that we have a \textasciitilde2.3 cm systematic uncertainty in our placement of the resistor. This corresponds to as much as a 27~$\mu$s difference in the $\Delta\,t$ between two sensors.

\begin{figure}[tbp]
    \centering
    \begin{subfigure}[b]{0.47\textwidth}
        \includegraphics[width=\textwidth]{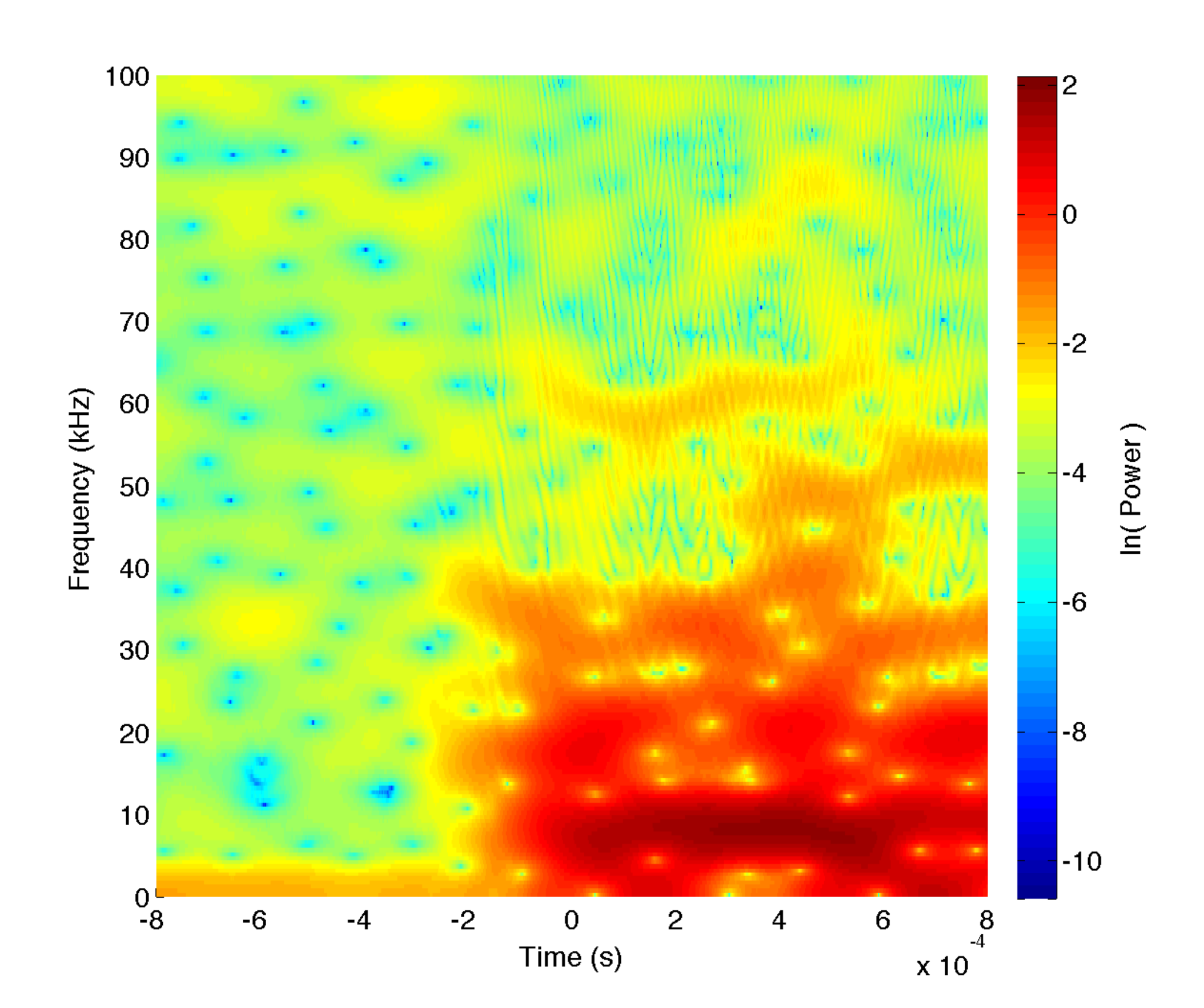}
        \caption{Bubble at center, z = 0 cm}
        \label{fig:BottomCenterSpec}
    \end{subfigure}
    ~ 
    \begin{subfigure}[b]{0.47\textwidth}
        \includegraphics[width=\textwidth]{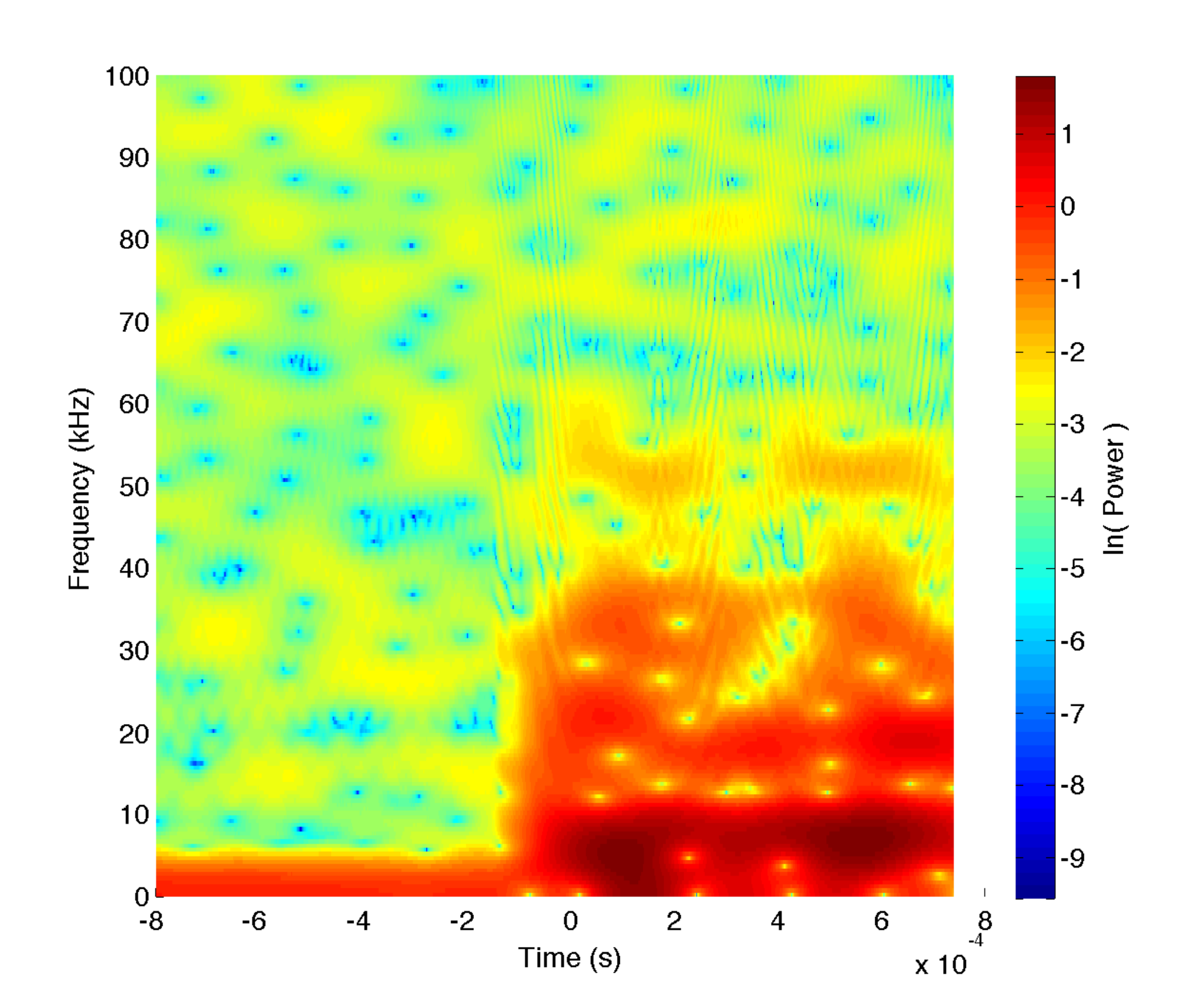}
        \caption{Bubble at center, z = 26.5 cm}
        \label{fig:TopCenterSpec}
    \end{subfigure}
        \caption{Spectrograms of the bubble. Frequency bands are visible at approximately 8~kHz, 18~kHz, and 30~kHz. The bubble directly against the bottom of the dewar shows early arriving sound in the higher frequency bands.  These frequencies can be filtered to isolate sound traveling primarily through the liquid nitrogen, as explained in the text.  }
        \label{fig:TransmissionSpectrograms}
\end{figure}
 
It is also valuable to understand the frequency components of sound traveling through the two media. We compute the discrete short-time Fourier transform of the of the signals, downsampled to 0.5MHz, using a Hamming window with a width of 150 samples. Figure \ref{fig:TransmissionSpectrograms} shows the resulting spectrograms for channel 1 of the events shown in Figure \ref{fig:TransmissionWaveforms}. The fundamental frequency of the bubble is evident near ~8 kHz.  In the bubble at the bottom, there is evidence of early sound at higher frequencies (\textasciitilde18 kHz and \textasciitilde30 kHz). These observations are consistent in all measurements at these locations. We interpret this as evidence that the steel preferentially transmits these higher frequencies.

From this analysis, we draw two conclusions. First, the sound transmission through the steel walls of the dewar is detectable, but has an uncertainty on the order of the time differences themselves which precludes its use in position reconstruction of bubbles. Second, this sound is preferentially at higher frequencies, so it may be possible to isolate sound transmission through the liquid nitrogen by constructing an appropriate band-pass filter.  We explore position reconstruction using frequency isolation in the following section.

\subsection{Position reconstruction using TDOA}

\paragraph{}The time difference of arrivals (TDOA) technique is a method for determining the position of a signal transmitter when the initial transmission time is unknown. It has been employed extensively in navigation and precise location of mobile devices.  For a fixed speed of sound, the time difference between two sensors receiving a signal defines a hyperboloid in 3D space on which the source can lie.  The problem then becomes finding the intersection of the surfaces defined by all the time differences observable with the receivers in the system. The exact solution for the 3D case is laid out in \cite{BucherMisra}.  However, in the presence of noise, it is shown in \cite{Gustafsson} that a best-fit approach performs better in the 2D case than the exact solution, and is much more easily implemented. We therefore generalize the latter to three dimensions and employ it here.

We create a least-squares cost function $J({\bf x}_b)$ that is minimized when the measured time difference between each pair of sensors, $\Delta\,t_{ij}$, is closest to the calculated time difference at a guessed bubble position, $\Delta\,t'_{ij}$. The calculation assumes that the sound travels in a straight line through the liquid nitrogen to the sensor.

\begin{equation}
J({\bf x}_b) = \sum_{i = 1}^4 \sum_{j=i}^4  (\Delta\,t_{ij} - \Delta\,t'_{ij})^2
\end{equation}

The position vector ${\bf x}_b = (x_b, y_b, z_b)$ represents the guessed location of the bubble at any given iteration of the minimization, and is related to the time difference of arrival between sensors $i$ and $j$ by the equation
\begin{equation}
\Delta\,t'_{ij}  = \frac{ ||{\bf x}_b - {\bf x}_i ||  - ||{\bf x}_b - {\bf x}_j || }{c_{LN}}
\end{equation}
where ${\bf x}_i$ and ${\bf x}_j$ are the positions of the sensors and the speed of sound in liquid nitrogen is taken to be $c_{LN} = 853$~m/s \cite{LNSpeed}. The position ${\bf x}_b$ floats in the minimization, subject to the constraint that it remain inside the cylindrical volume of the dewar.

To remove the high-frequency components, the waveforms are first downsampled to a sampling rate of 0.5~MHz.  We then employ two Chebyshev band-pass filtering schemes.  In the first, a four-pole high-pass filter is applied with a cutoff frequency $f_c = 12.5$~kHz to remove low-frequency components. This is followed by a four-pole low-pass filter with the same cutoff is applied to attenuate the high frequency components of the sound traveling through the steel walls.  In the second scheme, we utilize three two-pole filters in series: high-pass with $f_c = 5$~kHz, low-pass with $f_c = 2.5$~kHz, and another high-pass with $f_c = 5$~kHz. 

\begin{figure}[tbp]
    \centering
    ~ 
         \begin{subfigure}[b]{0.47\textwidth}
        \includegraphics[width=\textwidth]{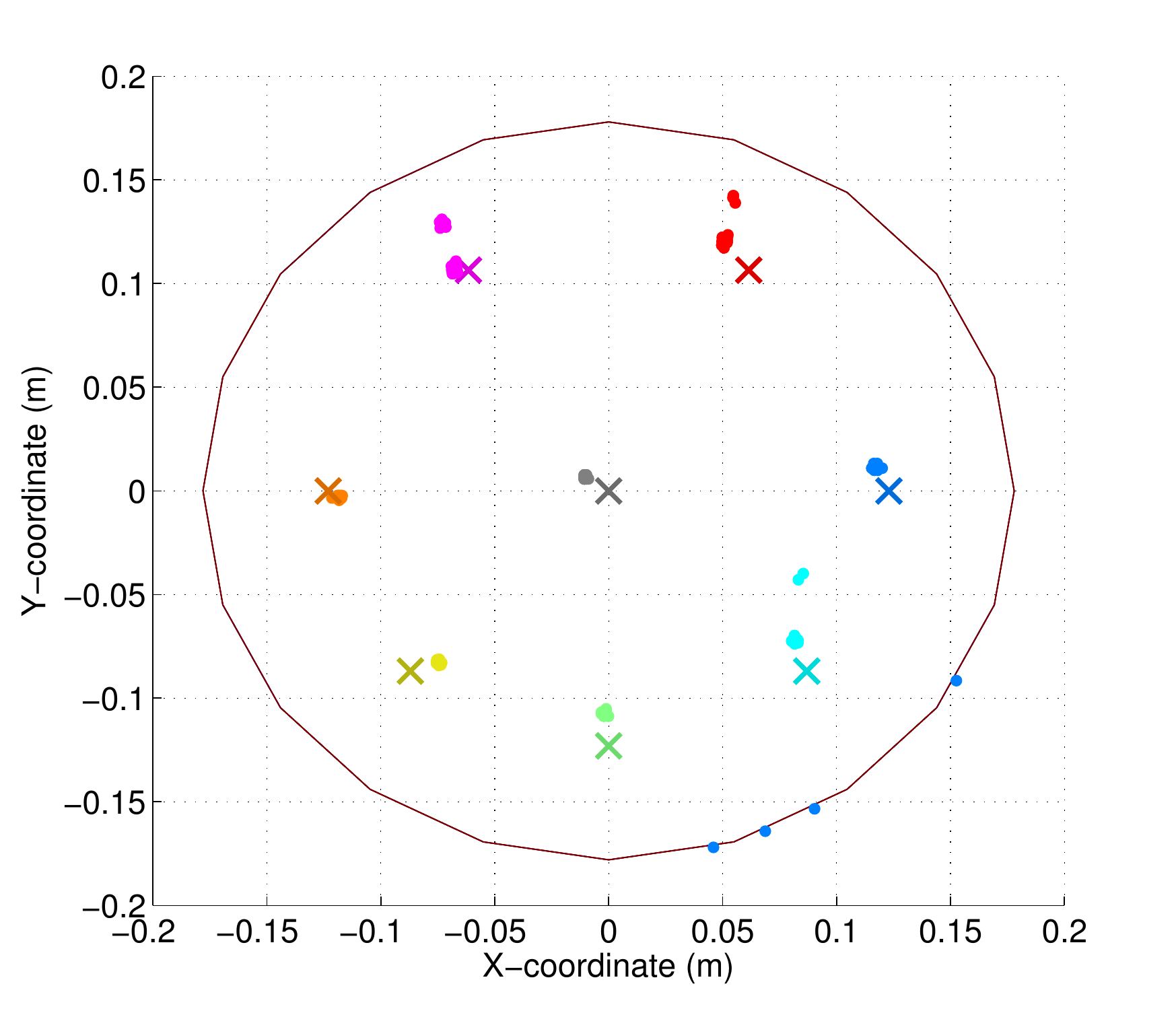}
        \caption{z=21cm}
        \label{fig:topXY}
     \end{subfigure}
    \begin{subfigure}[b]{0.47\textwidth}
        \includegraphics[width=\textwidth]{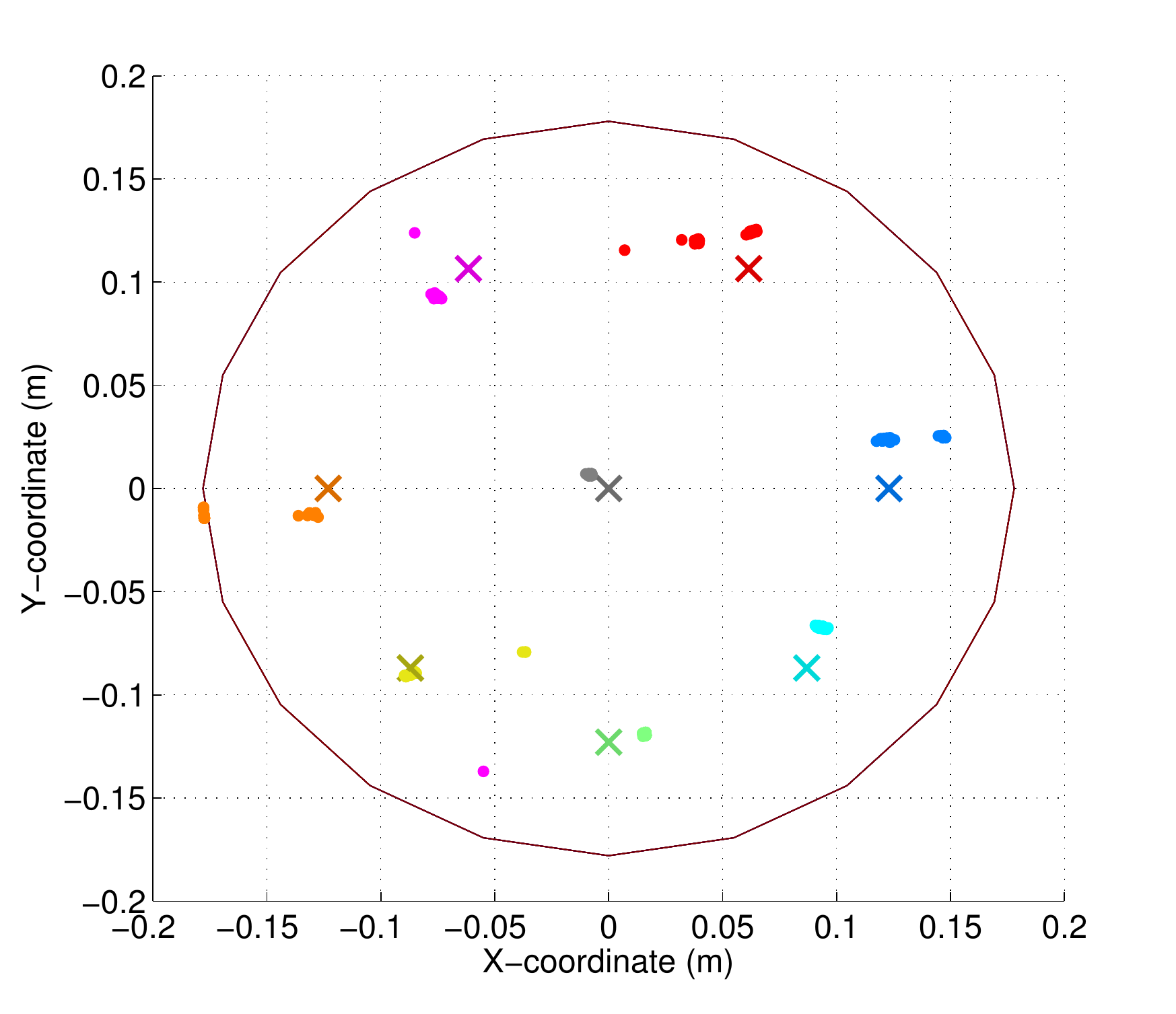}
        \caption{z = 10 cm}
        \label{fig:MiddleXY}
    \end{subfigure}
        \begin{subfigure}[b]{0.47\textwidth}
        \includegraphics[width=\textwidth]{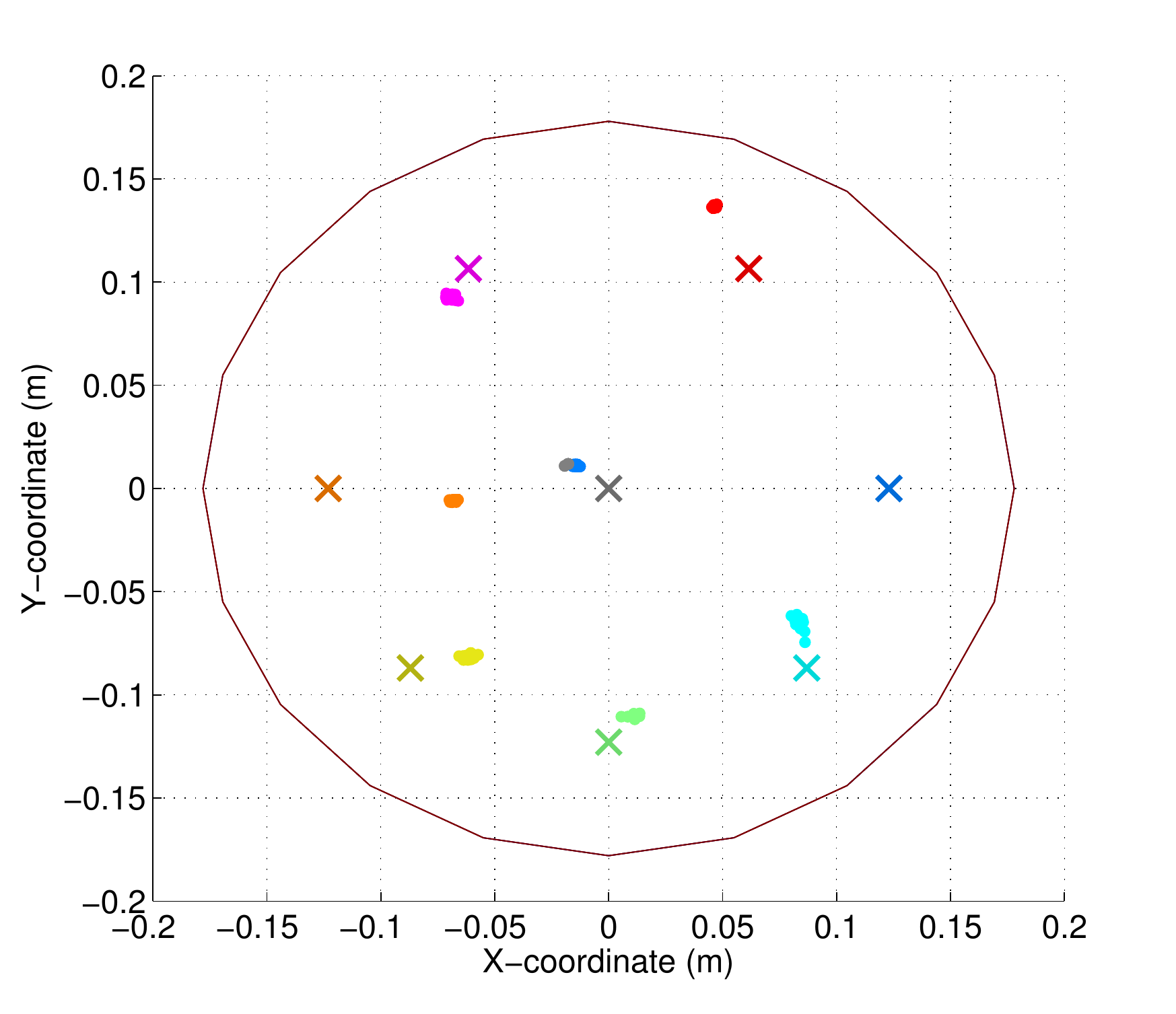}
        \caption{z = 0 cm}
        \label{fig:BottomXY}
    \end{subfigure}
     \begin{subfigure}[b]{0.47\textwidth}
       \includegraphics[width=\textwidth]{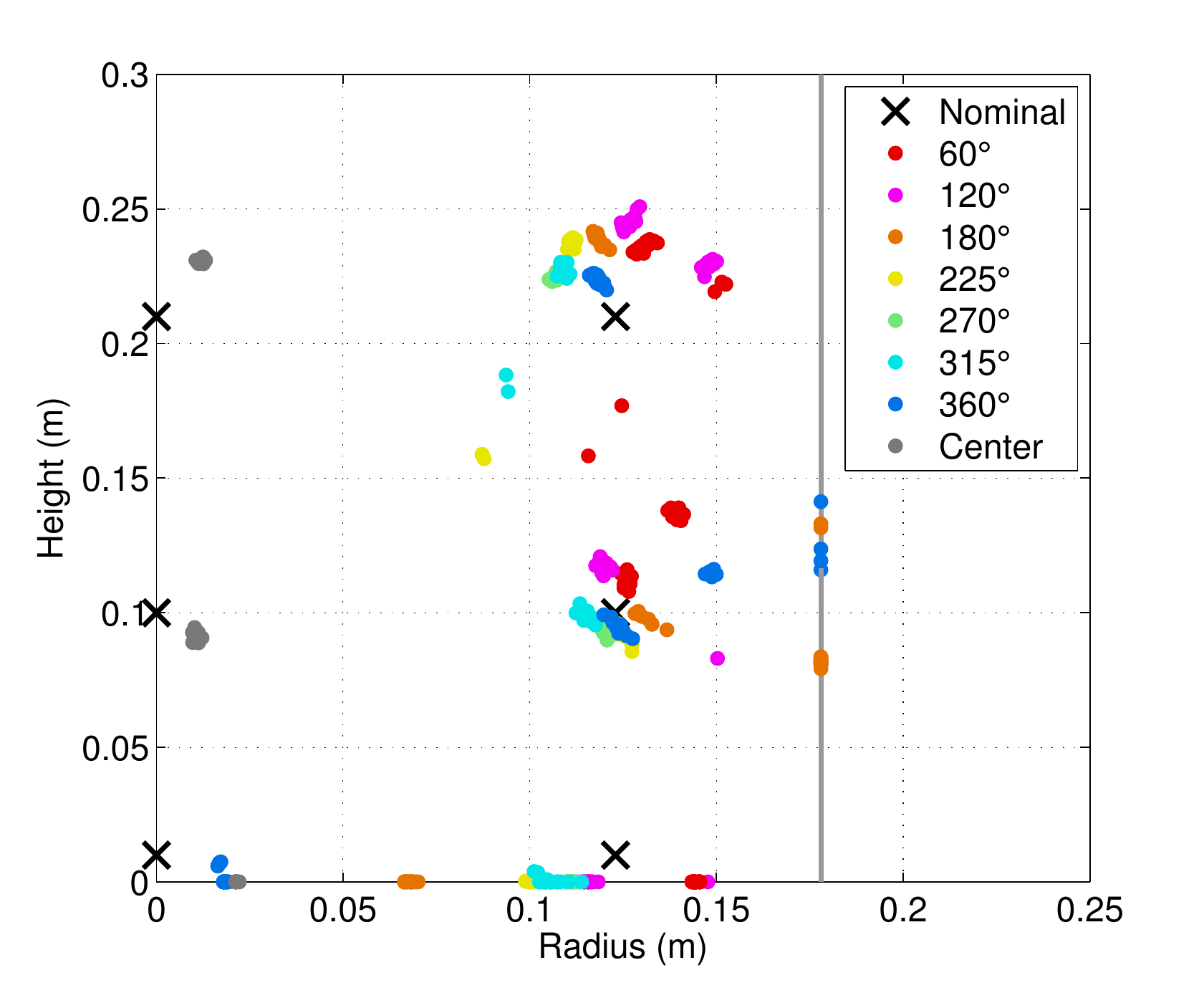}
       \caption{Height vs. radius}
       \label{fig:RZ}
       \end{subfigure}
        \caption{Radial, angular and position reconstruction of bubbles at different heights. Reconstructed clusters are self-consistent to within an average $\pm0.5$~cm, and are an average 2.5~cm from the nominal position of the resistor at each location. These systematic offsets are attributed primarily to a systematic uncertainty in the true position in the resistor. The first filtering scheme described in the text is used in (a) and (b), while the second scheme is used in (c). The reconstruction of height and radius is shown in (d), with the wall of the dewar drawn in grey at $r = 17.8$~cm. The nominal resistor position at each location is shown by the cross, while the reconstructed bubble positions are shown by circular points.}
        \label{fig:XYPositions}
\end{figure}

We test our position reconstruction on the data from June 2015. It is found that the first filtering scheme works well for bubbles in the bulk liquid. The reconstructed x-y positions are shown in Figures \ref{fig:topXY} and \ref{fig:MiddleXY}.  This scheme fails, however, for the measurements at the bottom of the dewar, and the second scheme is applied in Figure \ref{fig:BottomXY}. Close to the walls, it is likely that reflections from the nearby surface and a higher power of sound being transferred through the steel walls affect the frequency components picked up by the sensors, and necessitate a different filter. 

We observe very consistent reconstruction, with clusters having an average standard deviation of $< 0.5$cm is $x$, $y$, and $z$ after throwing away reconstructions that fit to the radial edges of the cylinder. Of the 707 events in the June dataset, 25 are reconstructed to the edges and 5 more are reconstructed farther than $3\,\sigma$ away from their mean position, indicating a convergence rate of $95.8\%$. Systematic offsets from the expected source positions (shown by the stars in Figure \ref{fig:XYPositions}) are observed in all cases.  The blue points at $z=0$ are the only group that we are confident are systematically reconstructed at the wrong position, and represent $33/707 =  4.7\%$ of the events.  The remaining clusters are an average 2.51~cm away from the nominal resistor positions, which can be partially explained by a systematic uncertainty in the true position of the bubble source. We estimate this uncertainty to be \textasciitilde$2.3$~cm by studying the reconstructed distance between similar resistor positions in the June and April datasets.



\section{Conclusion}
\paragraph{}We have presented here a demonstration of the use of piezoelectric sensors to detect and locate bubbles in a volume of cryogenic liquid. Our results indicate that bubbles can be detected and positioned reliably within the volume. Sources of bubbling in a cryogenic liquid detector could potentially be discovered and diagnosed using this technique.  We also demonstrate that sound  traveling through media other than the liquid is detected by the sensors, implying that an application in a working detector will need to appropriately model propagation and design signal filters to account for the effects of internal materials. 

To fully reconstruct a 3D position using TDOA information, at least four sensors are needed. However, it is likely that convergence and accuracy of the position reconstruction would be improved by additional sensors.  It may be possible to resolve serious systematic offsets by eliminating certain sensors from the measurement or utilizing all sensors to break degeneracies. 

In low-background  particle physics experiments, the piezoelectric sensors used in this work will not meet the stringent low-radioactivity requirements.  However, dark matter search experiments using superheated bubble chambers have successfully built and operated low-background acoustic sensors \cite{COUPP, PICO}. The techniques described in this work can be adapted to such a sensor for these applications.

\acknowledgments
The authors would like to acknowledge Marshall Styczinski and Gavin Fields for preliminary efforts on the present work. We would also like to thank Ray Gerhard, Britt Holbrook, David Hemer, and Keith DeLong for their engineering expertise and support.   The sensor readout circuit was designed by Ilan Levine of Indiana University South Bend. This work at the University of California, Davis was supported by U.S. Department of Energy grant DE-FG02-91ER40674,  as well as supported by DOE grant DE-NA0000979, which funds the seven universities involved in the Nuclear Science and Security Consortium. Brian Lenardo is supported by the Lawrence Scholars Program at the Lawrence Livermore National Laboratory (LLNL). LLNL is operated by Lawrence Livermore National Security, LLC, for the U.S. Department of Energy, National Nuclear Security Administration under Contract DE-AC52-07NA27344.

\end{document}